\documentclass[12pt,nohyper,letterpaper]{JHEP3}         
\usepackage{amssymb,amsfonts}

\usepackage{epsfig}

\def\be{\begin{eqnarray}}
\def\ee{\end{eqnarray}}
\newcommand{\nn}{\nonumber}
\newcommand\para{\paragraph{}}

\newcommand{\eqn}[1]{(\ref{#1})}

\def\Dslash{\,\,{\raise.15ex\hbox{/}\mkern-12mu D}}
\def\Dbarslash{\,\,{\raise.15ex\hbox{/}\mkern-12mu {\bar D}}}
\def\delslash{\,\,{\raise.15ex\hbox{/}\mkern-9mu \partial}}
\def\delbarslash{\,\,{\raise.15ex\hbox{/}\mkern-9mu {\bar\partial}}}
\def\pslash{\,\,{\raise.15ex\hbox{/}\mkern-9mu p}}
\def\calDslash{\,\,{\raise.15ex\hbox{/}\mkern-12mu {\cal D}}}

\newcommand{\Tr}{{\rm Tr}}

\def\lae{\mathrel{\mathop{\smash{\lower .5 ex \hbox{$\stackrel<\sim$}}}}}
\def\lae{\mathrel{\mathop{\smash{\lower .5 ex \hbox{$\stackrel>\sim$}}}}}

\def\nn{\nonumber}

\def\be{\begin{equation}}
\def\ee{\end{equation}}
\def\ba{\begin{array}}
\def\ea{\end{array}}

\def\dalemb#1#2{{\vbox{\hrule height .#2pt
        \hbox{\vrule width.#2pt height#1pt \kern#1pt
                \vrule width.#2pt}
        \hrule height.#2pt}}}

\newcommand{\bea}{\begin{eqnarray}}
\newcommand{\eea}{\end{eqnarray}}

\title{Instantons and Emergent AdS${}_3\times$S${}^3$ Geometry}
\author{Heng-Yu Chen and David Tong \\
Department of Applied Mathematics and Theoretical Physics, \\
University of Cambridge, UK\\
{\tt h.y.chen, d.tong@damtp.cam.ac.uk}}

\abstract{Two dimensional ${\cal N}=(4,4)$ gauge theories flow to
interacting superconformal field theories on their Higgs branch.
We examine  worldsheet instantons in these theories through the
eyes of a D-brane construction. The effective instanton partition
function is shown to reveal an emergent background $AdS_3\times
S^3$ geometry.
\para
{}
\para
\begin{center}
{\it Dedicated to the memory of John Brodie and Andrew 
Chamblin\\ and the roadtrip to Roswell.}\end{center}}

\begin{document}
\pagestyle{plain} \setcounter{page}{1}
\newcounter{bean}
\baselineskip16pt

\section*{1. Introduction}

The notion of gravity as an emergent force holds a certain appeal.
The no-go theorem of Weinberg and Witten \cite{ww} presents a
hurdle that any attempt to dynamically generate a generally
covariant theory of gravity must negotiate, and several examples
which evade the assumptions of the theorem are now known. The most
compelling of these is the celebrated AdS/CFT correspondence
\cite{juan} in which both gravity and the bulk of spacetime appear
dynamically.

\para
The emergence of a bulk geometry  from field theoretic origins is
a strong coupling phenomenon that is not easy to see. Indeed, it
is hard to determine from first principles which field theories
have a string dual.  A window is provided by BPS solitons. Within
the framework of the AdS${}_5$/CFT${}_4$ correspondence, it is
Yang-Mills instantons which provide a natural probe of the
emergent geometry where they are represented by D-instantons
which may roam the higher dimensional space
\cite{greenbank,colour}. It was shown by Dorey et. al.
\cite{dhkmv} that the partition function of instantons in $SU(N)$
${\cal N}=4$ super Yang-Mills coincides with that of the D-instanton. 
Most notably, the effective large $N$ moduli
space of Yang-Mills instantons in this theory is $AdS_5\times S^5$. The
radial direction of $AdS_5$ arises from the scaling mode of the
instanton while the internal $S^5$ space arises from
a correct treatment of the fermionic zero modes in the large $N$
limit.

\para
The techniques developed in \cite{dhkmv} were subsequently
extended to study the emergent geometry in other four-dimensional
gauge theories. These include ${\cal N}=2$ superconformal theories
\cite{n2}, orbifold theories \cite{orbi}, $SO(N)$ and $Sp(N)$
gauge groups \cite{spn1,spn,spn2}, field theories on their Coulomb
branch \cite{coulomb} and the $\beta$-deformation of the ${\cal
N}=4$ theory \cite{beta}.

\para
The purpose of this paper is to employ similar techniques in the
case of two-dimensional superconformal field theories (SCFTs). 
A geometry with an $AdS_3$ factor of radius $L$ is known to be dual to a 2d
conformal field theory with central charge $c=3L/2G_N$, with $G_N$ the 3d
Newton constant \cite{brownh}. The
most well studied examples are the ${\cal N}=(4,4)$ SCFTs arising
on the D1-D5 system wrapped on a 4-manifold $X\cong K3$ or ${\bf
T}^4$ for which the dual geometry is $AdS_3\times
S^3\times X$ (see \cite{review} for more details). The natural probes 
of the emergent geometry are the sigma-model instantons of the SCFT. Indeed, 
it was shown in \cite{mik} that these correspond to D-instantons roaming  
the six-dimensional supergravity background $AdS_3\times S^3$.

\para
Here we focus on simpler conformal field theories arising on the Higgs
branch of ${\cal N}=(4,4)$ gauge theories which do not have a
known AdS dual, in particular 2d SQCD with $N_f>N_c$. As we shall review, 
sigma-model instantons only arise when the
singularity of the Higgs branch is lifted through the introduction
of a Fayet-Iliopoulos term in the gauge theory. There are some
subtleties with sigma-model instantons which do not appear in the
case of four-dimensional Yang-Mills instantons. Among them is the
fact that the scaling mode suffers a
logarithmic divergence for conformal field theories living on the plane.
We evade this problem by working with the D-brane realization of
these instantons presented in \cite{amime}, for which
this divergence is cured. Our primary result is that, in the limit
of large central charge, the instantons naturally reside in an emergent
$AdS_3\times S^3$ geometry.

\para
The organization of the paper is as follows: Sections 2 and 3
contain review material on  ${\cal N}=(4,4)$ SCFTs on the Higgs 
branch and their instantons respectively.
The meat of the paper is in Section 4. Here we examine the
partition function of the D-brane realization of sigma-model
instantons and show how 
the effective instanton moduli space becomes $AdS_3\times S^3$.

\section*{2. The Conformal Field Theory of the Higgs Branch}
\setcounter{section}{2}

Aspects of ${\cal N}=(4,4)$ gauge theories in $d=1+1$ dimensions
were discussed in \cite{diacseib,wittenhiggs,ofer} and we restrict
ourselves to a brief summary of the relevant points.  These theories
can be thought of as the dimensional
reduction of ${\cal N}=(1,0)$ theories in six dimensions. The
$SU(2)_R$ R-symmetry that is enjoyed by all theories with eight
supercharges is enhanced upon dimensional reduction to
\be R=SU(2)_R\times SU(2)_l\times SU(2)_r\label{r}\ee
We work with gauge group $G$. The vector multiplet consists of a
gauge potential $A$,  four real scalars $V$ transforming under $R$
as $({\bf 1}, {\bf 2}, {\bf 2})$, together with four Weyl fermions.
There is also a triplet of auxiliary fields $D$ transforming as
$({\bf 3}, {\bf 1}, {\bf 1})$. The
hypermultiplet also contains four real scalars $H$ lying in the
$({\bf 2},{\bf 1},{\bf 1})\oplus ({\bf 2},{\bf
1},{\bf 1})$ representation and an additional four Weyl fermions.
The scalar potential of the theory is schematically
\be {\mathbb V}_{2d}=\frac{1}{2e^2}\Tr\,\left([V,V]^2 - D^2\right)
+ \sum_{\rm hypers} (V H)^2 + DHH\label{2dpot}\ee
where $e^2$ is the gauge coupling which, in $d=1+1$, has scaling
dimension $[e^2]=2$. Solutions to the classical vacuum equations
${\mathbb V}_{2d}=0$ fall roughly into two classes. The Coulomb
branch has $H=0$ and is parameterized by commuting matrices $V$.
In contrast, the Higgs branch has $V=0$ and is parameterized by
$H$ subject to gauge identification and the condition $D=0$. There
can also be mixed branches. Quantum mechanically the ground state
wavefunction spreads over these classical moduli spaces of vacua.

\para
In the far infra-red, a limit which requires $e^2\rightarrow
\infty$, our $d=1+1$ gauge theory is expected to flow to an
interacting ${\cal N}=(4,4)$ SCFT. In fact, it is expected to flow
to two decoupled superconformal field theories: one
arises on the Higgs branch and the other on the Coulomb branch.
Classically these two branches are connected at $V=H=0$, but
quantum effects can be shown to push this point to infinity in
field space through the generation of ``throat"-like wavefunction
renormalization \cite{diacseib,swd1d5,ofer}. The simplest argument
to show that the two branches indeed decouple uses the R-symmetry
in the ${\cal N}=(4,4)$ superconformal algebra \cite{wittenhiggs}
which includes both  left-moving and  right-moving $SU(2)$
R-symmetries. Since a symmetry which rotates scalar fields cannot
be split into left- and right-moving pieces, the $SU(2)_R$
symmetry of \eqn{r} cannot be part of the R-symmetry of the Higgs
branch SCFT, while the $SU(2)_l\times SU(2)_r$ R-symmetry cannot
be part of the R-symmetry of the Coulomb branch SCFT.

\para
In this paper we will focus on the conformal field theory of the
Higgs branch, for which the $SU(2)_l\times SU(2)_r$ R-symmetry of
the gauge theory happily descends to the SCFT. The central charge
of the theory coincides with the dimension of the Higgs branch:
$\hat{c}=2(n_H-n_V)$ where $n_H$ and $n_V$ are the number of
hypermultiplets and vector multiplets respectively. This central
charge can be confirmed by computing the anomaly in the two point
function of R-symmetry currents \cite{wittenhiggs}.

\para
The Higgs branch of \eqn{2dpot} has a singularity at $H=0$ where
gauge transformations do not act freely. If the gauge group $G$
includes abelian factors, there exists a set of marginal
deformations which resolve this singularity. From the perspective
of the gauge theory these arise through the addition of a triplet of
Fayet-Iliopoulos (FI) parameters $r$ transforming under the
R-symmetry as  $({\bf 3}, {\bf 1}, {\bf 1})$, together with a
theta term $\theta$,
\be {\rm Tr}_G\,(rD+i\theta F_{12})\ee
As we will review in section 3, the geometric resolution of the
singularities also introduces worldsheet instantons into the Higgs
branch theory in the guise of gauge theoretic
vortices\footnote{The metric on the Higgs branch does not receive
quantum corrections, suggesting that the singular point $H=V=0$
may lie at finite distance in field space. Nonetheless, there is
another description of the physics near the singularity which
exhibits a ``throat"-like structure \cite{swd1d5,ofer} with a
continuous spectrum of dimensions for primary operators.  In this
description, the marginal deformation which resolves the
singularity gives rise to a Liouville potential. It would be
interesting to study the role of instantons in the throat.}.
Before turning to the instantons, we first describe the specific 
${\cal N}=(4,4)$ gauge theory of interest: 2d SQCD.

\subsubsection*{2d SQCD}

Our primary focus will be on 2d SQCD with gauge group $G=U(N_c)$ and
$N_f$ hypermultiplets transforming in the fundamental representation
of $G$. The theory exhibits an $SU(N_f)$ flavor symmetry. We write the
hypermultiplets $H$ as a doublet of complex scalars, $H=(Q,\tilde{Q}^\dagger)$,
so that $Q$ transforms as $({\bf N}_c, \bar{\bf N}_f)$ and $\tilde{Q}$ as
$(\bar{\bf N}_c, {\bf N}_f)$ under $G\times SU(N_f)$. After a suitable $SU(2)_R$
rotation of the FI parameters, the triplet of D-term equations $D=0$
decomposes into the F-flatness condition
\be \sum_{a=1}^{N_f}Q_a\tilde{Q}_a=0 \label{f}\ee
together with the D-flatness condition
\be \sum_{a=1}^{N_f}Q_aQ_a^\dagger-\tilde{Q}_a^\dagger\tilde{Q}_a=r\label{d}\ee
where we have explicitly displayed the flavor index $a=1,\ldots, N_f$,
while both \eqn{f} and \eqn{d} are adjoint valued in $G$. The Higgs branch
${\cal M}$ is defined
as the solutions to these equations modulo the gauge action $Q\rightarrow UQ$,
$\tilde{Q}\rightarrow \tilde{Q}U$,  with $U\in G$. Solutions to \eqn{d} only
exist for $N_f\geq N_c$ and the complex dimension of the Higgs branch is
given by
\be \hat{c}=2N_c(N_f-N_c) \ee
The Higgs branch is the cotangent bundle
of the Grassmannian:  ${\cal M}\cong T^\star G(N_c, N_f)$. The zero section
$G(N_c,N_f)$, with radius $\sqrt{r}$, lies at $\tilde{Q}=0$.

\para
Since the Higgs branch is known to flow to an interacting
conformal field theory, one may wonder whether there is a gravity
dual in the limit of large central charge $\hat{c}$. None is
known. However, 2d SQCD may be realized on the worldvolume of
D-branes via the usual Hanany-Witten set-up, suspending $N_c$
D2-branes between two NS5-branes in type IIA string theory with
the role of the flavors is played by $N_f$ D4-branes
\cite{brodie,mohsen}. Our main result in Section 4 is that a
suitable probe of the IIA brane configuration does indeed reveal
an $AdS_3\times S^3$ geometry that one may expect for the gravity
dual of an ${\cal N}=(4,4)$ SCFT.

%
%
%
%
%
%
%

\section*{3. Worldsheet Instantons on the Higgs Branch}
\setcounter{section}{3}

Turning on the FI parameter $r$ resolves the singularity on the
Higgs branch by blowing up a two-cycle. This introduces worldsheet
instantons into the conformal field theory on the Higgs branch
which wrap this two-cycle. From the perspective of the gauge
theory these instantons arise as Nielsen-Olesen vortices
\cite{wittphases} (sometimes referred to in the soliton literature
as ``semi-local'' vortices \cite{semi,schroers}). In what follows,
we shall use the word ``instanton'' and ``vortex''
interchangeably.

\para
The instanton number is labelled by the first Chern class, the
integral of the magnetic flux $F_{12}$ over the Euclidean plane,
\be -\frac{1}{2\pi}\int\,\Tr\,F_{12}=k\in{\bf Z}\ee
In a semiclassical calculation, one must integrate over the Higgs
branch weighted by the ground state wavefunction. However, from the
perspective of  vortices, not all classical vacua on the Higgs
branch are created equal: BPS vortices, which descend to
holomorphic sigma model instantons in the infra-red limit, exist
only on the submanifold of the Higgs branch which emerges from the
singularity at finite FI parameter $r$. In the specific case of
SQCD described in the previous section, this means that BPS
vortices only exist on the $G(N_c,N_f)$ zero section of the Higgs
branch, defined by $\tilde{Q}=0$. These vacua are distinguished by
the spontaneous symmetry breaking of the gauge and flavor groups:
\be U(N_c)\times SU(N_f)\rightarrow[U(N_c)_{\rm diag}\times
U(N_f-N_c)]/U(1)\label{breaking}\ee
For $k>0$, the first order vortex equations require
$\tilde{Q}=V=0$ and
\be F_{12} =e^2 (\sum_{a=1}^{N_f}Q_aQ_a^\dagger -r)\ \ \ ,\ \
\ {\cal D}_zQ_a=0\label{vort}\ee
with $z=x^1+ix^2$. Vortices with $k<0$ are related by a parity
transformation. Bosonic solutions to \eqn{vort} have action
\be
S_{\rm inst} = 2\pi(r+i\theta)k
\label{action}\ee
We denote the moduli space of solutions to the vortex equations as
${\cal V}$. The moduli space has real dimension ${\rm dim}({\cal
V})=2kN_f$. For $k=1$ and $N_f=N_c$ --- the case where there is no
Higgs branch --- there are two translational zero modes dictating
the position of the vortex in the plane. A further $2(N_c-1)$
orientational modes specify how the vortex sits in surviving
$U(N_c)_{\rm diag}$ group of \eqn{breaking} \cite{amime,auzzi}.
All these modes have finite norm. The vortices are
exponentially localized objects, with a  fixed size $\rho\sim
1/e\sqrt{r}$. In the infra-red $e^2\rightarrow \infty$ limit they
become singular, point-like objects.

\para
When $N_f>N_c$, a true Higgs branch exists and the vortices have a
qualitatively different nature. The massless fields of the Higgs
branch may now be excited around the background of the vortex
resulting in a power-law tail specified by a size modulus $\rho
\geq 1/e\sqrt{r}$. In the limit $e^2\rightarrow\infty$, the
vortices no longer become singular, but descend to the sigma model
instantons on the Higgs branch. In this limit, the scaling modulus
of the instanton $\rho$ arises as a Goldstone mode from broken
conformal invariance. A further $2(N_f-N_c)-1$ orientational modes
for these massless fields arise as Goldstone modes of the
$U(N_f-N_c)$ factor of \eqn{breaking}.

\para
The extra modes of the semi-local vortex bring with them an extra
problem: they are non-normalizable for theories defined on ${\bf
R}^2$. This well known fact about sigma-model instantons
\cite{ward} is not ameliorated by the transition to finite $e^2$
\cite{leese,sy}. (Indeed, turning on finite $e^2$ changes the
vortex solution only at short distance, while the divergent norm
is an infra-red issue). There are a number of ways to regulate
this divergence:

\begin{itemize}
\item One may define the field theory on the Riemann sphere with
radius $R$. The normalization of the scaling and orientation modes
now has an overall factor $\sim \log(R/\rho)$. It would certainly
be interesting to reconsider the calculation of Section 4 in this
case.

\item One may introduce distinct masses $m$ for the fundamental
hypermultiplets. Classically this lifts the Higgs branch. It also
lifts the scaling and orientation modes, endowing  them with a
mass $m$ and regulating their norm as $\sim\log(1/m\rho)$
\cite{sy}. The effect on the sigma-model instantons is similar to
the effect of moving onto the Coulomb branch on a Yang-Mills
instanton: it causes the instanton to shrink to zero size. A
computation of the emergent geometry of the Coulomb branch of
${\cal N}=4$ super Yang-Mills was performed in \cite{coulomb}.

\item One may realize the sigma-model instantons as D-branes in the
Hanany-Witten type set-up \cite{amime}. While the moduli space ${\cal V}$
of the instanton and D-brane coincide \cite{amime} (see \cite{sy} for a 
recent discussion specifically addressing issues for semi-local vortices), 
the metrics differ and, most notably, the logarithmic 
divergences are rendered finite.
\end{itemize}
In this paper we focus on the latter of these regulators and show
how the D-brane realization of sigma-model instantons sees an 
emergent background geometry. This has the disadvantage that the
computation is not purely field theoretic: we are working in the
D-brane regime, rather than the sigma-model limit. The advantage
is that the D-brane realization of the instanton partition
function takes a convenient and familiar form as we now review.

\subsubsection*{A D-brane Realization of Instantons on the Higgs
Branch}

\EPSFIGURE{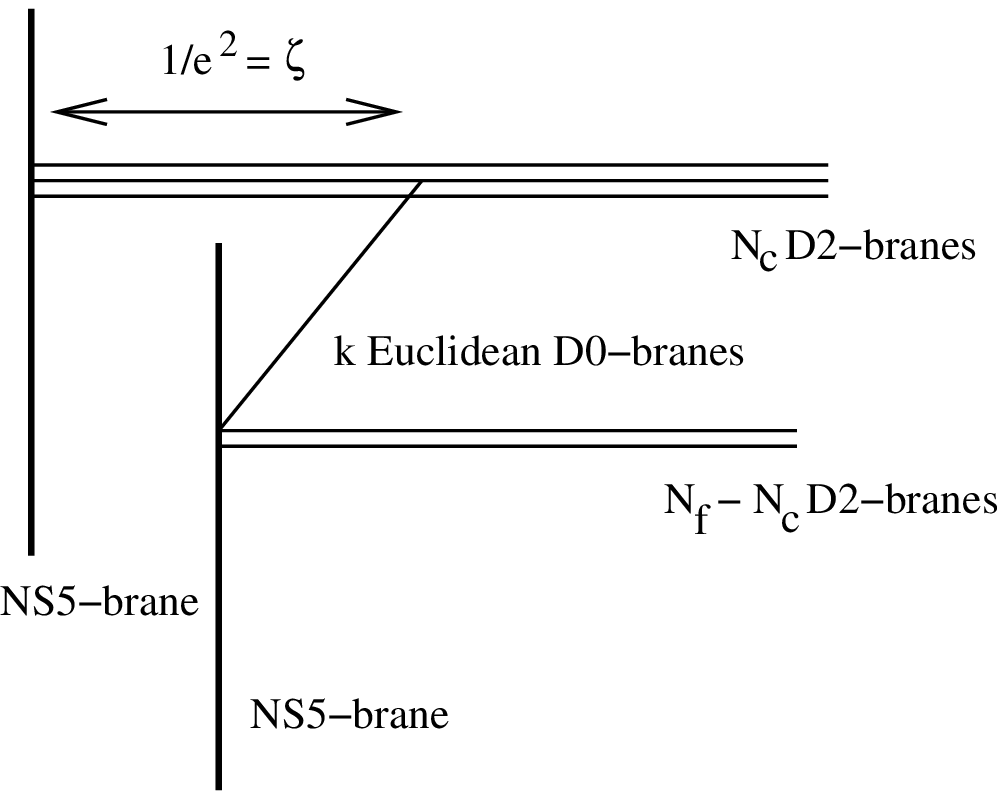,height=140pt}{}
A D-brane realization of vortices in eight supercharge SQCD was
presented in \cite{amime} using Hanany-Witten type brane
configurations. In the present context, this consists of D2-branes
suspended between parallel NS5-branes, with further, semi-infinite D2-branes 
providing the hypermultiplets as shown in the figure \cite{brodie,mohsen}. 
Separating the NS5-branes in
the direction out of the page induces the FI parameter, forcing the
2d gauge theory onto the Higgs branch. The instantons are stretched
Euclidean D0-branes as depicted.

\para
The worldvolume theory on the D0-branes is a $d=0+0$ matrix model
with ${\cal N}=(2,2)$ supersymmetry (i.e. the complete dimensional
reduction of theory with ${\cal N}=1$ supersymmetry in $d=3+1$). We
refer to this as the ``vortex matrix model''; it
consists of the following fields \cite{amime}

\begin{itemize}
\item $U(k)$ vector multiplet, containing four real scalars $X_a$,
$a=1,2,3,4$ and a doublet of complex Grassmannian parameters
$\lambda_\alpha$, $\alpha=1,2$. \item $N_c$ chiral multiplets
transforming in the fundamental ${\bf k}$ representation, each
containing complex scalars $\phi_i$, $i=1,\ldots, N_c$ and $2N_c$
complex fermions $\mu_{i\alpha}$. \item $N_f-N_c$ chiral
multiplets transforming in the anti-fundamental $\bar{\bf k}$
representation, containing complex scalars $\tilde{\phi}_r$,
$r=1,\ldots,N_f-N_c$ and $2(N_f-N_c)$ complex fermions
${\tilde{\mu}}_{i\alpha}$. \item A single chiral multiplet
transforming in the adjoint representation of $U(k)$. This
contains a complex scalar $a$, together with a complex
doublet of fermions $\chi_\alpha$.
\end{itemize}
The gauge coupling constant $g^2$ of the vortex matrix model is
taken to infinity. The only parameter in the theory is a FI
parameter $\zeta$, inherited from the 2d gauge coupling $e^2$,
\be \zeta= \frac{2\pi}{e^2}\ee
The full action will be presented shortly in its full indexical
glory. Here we restrict attention to the $U(k)$ valued D-flatness
condition which, since we work in the limit
$g^2\rightarrow\infty$, is strictly imposed in the matrix model
\be [a,a^\dagger]^m_{\
n}+\sum_{i=1}^{N_c}\phi^m_i\phi_{n}^{\dagger\,i}
-\sum_{r=1}^{N_f-N_c}
\tilde{\phi}_r^{m\dagger}\tilde{\phi}_{n}^r=\zeta\delta^m_{\ n}
\label{higgs}\ee
where $m,n=1,\ldots,k$ are gauge indices.  Solutions to
\eqn{higgs}, modulo $U(k)$ gauge transformations, define the
vacuum manifold ${\cal V}$ of the vortex matrix model. We have
${\rm dim}({\cal V})=2kN_f$ and ${\cal V}$ is identified with the
vortex moduli space \cite{amime}. However, as promised, the metric
on the Higgs branch ${\cal V}$, inherited through the usual
K\"ahler quotient construction from the flat metric on ${\bf
C}^{k(N_f+k)}$, does not suffer a divergent norm. Despite the 
differences in the metric, the vortex theory is expected to capture 
the BPS properties of field theoretic vortices, an assertion which has passed a 
number of checks \cite{vstring, tit, sy}.

\para
For finite $e^2$ we have $\zeta\neq 0$ and the $U(k)$ gauge
transformations are freely acting on solutions of \eqn{higgs},
ensuring that ${\cal V}$ is smooth as befits the vortex moduli
space. However, in the $e^2\rightarrow \infty$  infra-red limit of
the 2d gauge theory, we have $\zeta=0$ and ${\cal V}$ develops a
singularity. This reflects the existence of singular (``small'')
$\rho\rightarrow 0$ instantons in the 2d sigma model. In the
following we shall be interested in the infra-red SCFT on the 2d Higgs
branch, and correspondingly focus on the vortex theory with
$\zeta=0$.

\section*{4. The Instanton Partition Function}
\setcounter{section}{4}

In this section we study the vortex matrix model in more detail,
following closely the calculation of \cite{dhkmv}. The cultured
reader may notice the similarity between the vortex theory
described above and the ADHM matrix model of Yang-Mills
instantons. Indeed, the vortex moduli space can be shown to be a
complex submanifold of the moduli space of $SU(N_f)$ Yang-Mills
instantons \cite{amime}. This correspondence motivated the present
work and allows us to transfer the techniques developed in
\cite{dhkmv} for four-dimensional gauge theories to the present
case of two-dimensional sigma-models.

\para
The full action of the vortex matrix model can be written as the
sum of three terms,
\begin{eqnarray}
S_{\Phi}&=&\sum_{i=1}^{N_c}\left(\phi^{\dag\,i}X_{a}X_{a}\phi_i+\bar{\mu}^i_{\dot{\alpha}}
X_{a}\bar{\sigma}_{a}^{\dot{\alpha}\alpha}\mu_{i\alpha}
+i\sqrt{2}\phi^{\dag\,i}\lambda^{\alpha}\mu_{i\alpha}
-i\sqrt{2}\bar{\mu}^i_{\dot{\alpha}}\bar{\lambda}^{\dot{\alpha}}\phi_i-\phi^{\dag\,i}D\phi_i
\right) \nn\\
S_{\tilde{\Phi}}&=&\sum_{r=1}^{N_f-N_c}\left(\tilde{\phi}^rX_{a}X_{a}
\tilde{\phi}_r^{\dag}+
{\tilde{\mu}}^{r\alpha}X_{a}\sigma^{a}_{\alpha\dot{\alpha}}\bar{{\tilde{\mu}}}_r^{\dot{\alpha}}
-i\sqrt{2}{\tilde{\mu}}^{r\alpha}\lambda_{\alpha}\tilde{\phi}_r^{\dag}
+i\sqrt{2}\tilde{\phi}^r\bar{\lambda}_{\dot{\alpha}}\bar{{\tilde{\mu}}}_r^{\dot{\alpha}}
+\tilde{\phi}^rD\tilde{\phi}_r^{\dag}\right)\nn\\ S_{A}&=& {\rm
Tr}_{k} \left(-|[X_{a},a^{\dag}]|^2-
[X_{a},\bar{\chi}_{\dot{\alpha}}]\bar{\sigma}^{\dot{\alpha}\alpha}_{a}\chi_{\alpha}
+i\sqrt{2}\bar{\chi}_{\dot{\alpha}}[a,\bar{\lambda}^{\dot{\alpha}}]
+i\sqrt{2}[a^{\dag},\lambda^{\alpha}]\chi_{\alpha}+D[a,a^\dagger]
\right)\nn
\end{eqnarray}
Here only $U(k)$ gauge indices have been left implicit. All others
indices are explicitly displayed to expose the large global
symmetry group $F$
\be F= U(1)_R\times U(1)_E\times Spin(4)_R\times [U(N_c)\times
U(N_f-N_c)]/U(1)\label{vortsym}\ee
Of these, $U(1)_E$ acts only on the adjoint chiral multiplet
fields $a$ and $\chi$, and arises from Euclidean rotations of the
instanton on the plane ${\bf R}^2$. The non-abelian $[U(N_c)\times
U(N_f-N_c)]/U(1)$ factor is identified with the corresponding
factor in \eqn{breaking}, with the quotient $U(1)$ part of the
$U(k)$ gauge group. Finally, the $U(1)_R\times Spin(4)_R\cong 
U(1)_R\times SU(2)_l\times SU(2)_r$ is identified with the 2d R-symmetry 
\eqn{r}, where $SU(2)_R\rightarrow U(1)_R$ through the introduction of the 
FI parameter. The transformation of the various fields under
the $U(k)$ gauge group and the various flavor groups is shown in
the table. 
\begin{center}
\begin{tabular}{|c|c|c|c|c|c|c|c|}
\hline
 & $U(k)$&$U(1)_R$&$U(1)_E$&Spin(4)&$U(N_c)$& $U(N_f-N_c)$ \\
\hline \ \ $X$\ \ & {\bf adj} & 0 & 0 & ({\bf 2},{\bf 2}) & {\bf
1} & {\bf
1} \\ \hline $\lambda$ & {\bf adj} & +1 & 0 &({\bf 1},{\bf 2}) & {\bf 1} & {\bf 1} \\
\hline $\phi$ & {\bf k} & +2 & 0 & {\bf 1} & $\bar{\bf N}_c$ & {\bf 1} \\
\hline $\mu$ & {\bf k} & +1 & 0 & ({\bf 1},{\bf 2}) & $\bar{\bf N}_c$ & {\bf 1} \\
\hline $\tilde{\phi}$ & $\bar{\bf k}$ & +2 & 0 & {\bf 1} & {\bf 1} & ${\bf N_f-N_c}$ \\
\hline $\tilde{\mu}$ & $\bar{\bf k}$ & +1 & 0 & ({\bf 1},{\bf 2}) & {\bf 1} & ${\bf N_f-N_c}$ \\
\hline $a$ & {\bf  adj} & 0 & +1 & {\bf 1} & {\bf 1} & {\bf 1} \\
\hline $\chi$ & {\bf  adj} & -1 & +1 & ({\bf 1},{\bf 2}) & {\bf 1} & {\bf 1} \\
\hline
\end{tabular}
\end{center}
We have included here only the unbarred fermions; the barred 
fermions transform in the 
$({\bf 2},{\bf 1})$ representation of $Spin(4)_R$ and in the 
conjugate representation of the other symmetry groups. 

\para 
We are interested in evaluating the instanton partition
function
\be {\mathcal{Z}}_{k,N_{c},N_{f}}=\frac{1}{{\rm Vol}(U(k))}\int\,
\exp-\left(S_{\Phi}+S_{\tilde{\Phi}}+S_A\right)\,, \label{path}\ee
where the integration is over all fields listed in the table,
together with their complex conjugates as well as the auxiliary
field $D$. Integrating over the latter gives the D-flatness
condition \eqn{higgs}. Note that we have also stripped off the 
obvious constant factor of $\exp(-S_{\rm inst})$ given in \eqn{action}.

\para
The instanton contribution to a given correlator $\langle {\cal
O}\rangle$ is evaluated by inserting ${\cal O}_{\rm inst}$ into
the partition function ${\cal Z}_{k,N_c,N_f}$ where ${\cal O}_{\rm
inst}$ is the operator ${\cal O}$ evaluated on the instanton
background. Since ${\cal O}_{\rm inst}$ depends on the collective
coordinates of the instanton (i.e. the various fields of the
vortex matrix model) we obviously don't wish to perform all
integrations in ${\cal Z}_{k,N_c,N_f}$ before inserting ${\cal
O}$; rather we will perform a subset of the integrations to yield
an effective partition function which can subsequently be used to
evaluate correlators which are both gauge and flavor
singlets\footnote{This differs from the 4d calculation of
\cite{dhkmv}, where only the requirement of gauge invariance is
needed. It's worth noting however that in the most studied example
of $AdS_3/CFT_2$, the would-be flavor symmetry on the D5-brane
becomes gauged once it is compactified on $X={\bf T}^4$ or $K3$.}.
As we shall see, our effective partition function will be valid in
the limit of large $N_c$ and $N_f-N_c$, and hence large central
charge $\hat{c}$.

\para
Let us start by introducing singlet fields under the $U(N_c)$ and
$U(N_f-N_c)$ global symmetries of the matrix model. We define
\be W^{m}_{n}=\sum_{i=1}^{N_{c}}\phi^{m}_{i}\phi^{\dag\,i}_{n} \ \
\ \ ,\ \ \ \
\tilde{W}^{m}_{n}=\sum_{r=1}^{N_f-N_{c}}\tilde{\phi}^{\dag\,m}_{r}\tilde{\phi}^{r}_{n}
\label{ww}\ee
We are trading $2kN_c$ $\phi$ variables for $2k^2$ $W$ variables
(and similarly for $\tilde{\phi})$. When $k\geq N_c$, this isn't
an improvement; generically orientated instantons fill the entire
$SU(N_c)$ group space and we should stick with the $\phi$
variables. In contrast, when $k<N_c$, the instantons rattle around
inside the group space, filling at most a $SU(N_c-k)$ subgroup. In
this situation, the partition function picks up a volume factor
from the integration over the $SU(N_c)/SU(N_c-k)$ coset space. The
Jacobian factors that arise from changing from $\phi,
\tilde{\phi}$ to $W,\tilde{W}$ variables were computed in
\cite{dhkmv}, yielding
\be
\int\,d\phi d\tilde{\phi} = C_{k,N_c}C_{k,N_f-N_c} \int\,dW
d\tilde{W}\
\left( {\rm det}_{k}W\right)^{N_{c}-k} ({\rm det}_{k} \tilde{W})^{N_{f}-N_{c}-k}\,.
\ee
where the coefficients, which arise from the volume of the coset
spaces, are given by
\be C_{k,N} =
\frac{(2\pi)^{Nk-k(k-1)/2}}{\prod_{i=1}^{k}(N-i)!}\label{ckn}\ee
We may eliminate one of these integrals by imposing the D-flatness
constraint \eqn{higgs} which is now linear in the $W,\tilde{W}$
variables, reading: $W-\tilde{W}=[a,a^\dagger]$.

\para
We now turn to the fermions. The action
$S=S_\Phi+S_{\tilde{\Phi}}+S_A$ contains no quadratic terms for
the vector multiplet fermions $\lambda$, ensuring that they each
act as a Grassmannian Lagrange multiplier, imposing constraints on
the remaining fields:
\be [\chi^{\alpha},a^{\dag}]^{m}_{\ n}+\sum_{i=1}^{N_{c}}
\mu^{\alpha m}_{i}\phi^{\dag\,i}_{n}
-\sum_{r=1}^{N_{f}-N_{c}}\tilde{\phi}^{\dag\,m}_{r}
{\tilde{\mu}}^{\alpha r}_{n}=0
\label{fcons}\ee
Solutions to the complex constraint equations \eqn{fcons}, modulo
$U(k)$ gauge transformations, describe the physical fermionic zero
modes of the instanton. We wish to mimic the procedure just
performed in the bosonic sector, integrating out the superpartners
of the gauge and flavor orientation modes. Schematically these are
the fermions $\mu$ and $\tilde{\mu}$  satisfying
$\phi\bar{\mu}=\tilde{\phi}^\dagger\tilde{\mu}=0$. More precisely,
we may decompose the fermionic variables as
\be \mu^{\alpha m}_{i}=\theta^{\alpha m}_{\
n}\phi^{n}_{i}+\nu^{\alpha m}_{i}\ \ \ \ \ ,\ \ \ \ \
{\tilde{\mu}}^{\alpha r}_{m}=\tilde{\phi}^{r}_{\
n}\tilde{\theta}^{\alpha n}_{\ m}+\tilde{\nu}^{\alpha r}_{m}
\label{decomp}\ee
where $\nu$ and $\tilde{\nu}$ are the superpartners of the
orientation modes, satisfying
\be \sum_{i=1}^{N_c}\phi^{m}_{i}\bar{\nu}^{\dot{\alpha i}}_{n}=0\
\ \ \ \ {\rm and} \ \ \ \ \ \sum_{r=1}^{N_f}
\tilde{\phi}^{\dag\,m}_{r}\tilde{\nu}^{\alpha r}_{n}=0 \ee
In terms of these new variables, the constraints \eqn{fcons} can
be written in a manifestly $U(N_c)\times U(N_f-N_c)$ invariant
manner as
\be [\chi^{\alpha},a^{\dag}]^{m}_{\ n}+\theta^{\alpha m}_{\
l}W^l_{\ n} -\tilde{W}^m_{\ l}\tilde{\theta}^{\alpha l}_{\ n}=0
\label{newfcons}\ee
The change from Grassmann
coordinates $\mu$ and $\tilde{\mu}$ to $\theta,\tilde{\theta},\nu$
and $\tilde{\nu}$ entails a Jacobian factor, once again computed
in \cite{dhkmv}
\be\int\,d\mu\,\,d\tilde{\mu} =
\int\,d\theta\,d\tilde{\theta}\,d\nu\,d\tilde{\nu}\,
|\,{\rm{det}}_{k}W|^{-k}|\,{\rm{det}}_{k}\tilde{W}|^{-k}\,,\nonumber\\
\ee
To integrate over the Grassmannian super-orientation modes $\nu$
and $\tilde{\nu}$ we must see how they appear in the matrix model
action. In instanton computations, extraneous fermionic zero modes
are often lifted by a four-fermi coupling in the action. In the
present case such a coupling comes from integrating out the four
vector multiplet scalars $X$. The relevant terms in the vortex
matrix model are
\be
S_{X}=-{\mathrm{Tr}}_{k}\left(X_{a}{\mathbb{L}}[X_{a}]+X_{a}\Lambda_{a}\right)\,,
\label{X}\ee
where the linear operator $\mathbb{L}$ is given by
\be {\mathbb{L}}[X_{a}]^{m}_{\ n}=[a^{\dag},[a,X_{a}]]^{m}_{\ n}
+\frac{1}{2}\{X_{a},W+\tilde{W}\}^{m}_{\ n}\label{L} \ee
while the adjoint valued fermion bi-linear $\Lambda$ is
(neglecting gauge indices)
\bea \Lambda^{a}&=&\bar{\chi}_{\dot{\alpha}}
(\bar{\sigma}^{a})^{\dot{\alpha}\alpha} \chi_{\alpha}
+\chi^{\alpha}(\sigma^{a})_{\alpha\dot{\alpha}}
\bar{\chi}^{\dot{\alpha}}-\sum_{i=1}^{N_{c}}
\mu^{\alpha}_{i}(\sigma^{a})_{\alpha\dot{\alpha}}
\bar{\mu}^{\dot{\alpha}i}
-\sum_{r=1}^{N_{f}-N_{c}}\bar{{\tilde{\mu}}}_{\dot{\alpha}r}
(\bar{\sigma}^{a})^{\dot{\alpha}\alpha}{\tilde{\mu}}_{\alpha r}
\nn\\ &\equiv & \hat{\Lambda}^a -\sum_{i=1}^{N_{c}}
\nu^{\alpha}_i(\sigma^{a})_{\alpha\dot{\alpha}}
\bar{\nu}^{\dot{\alpha}i}
-\sum_{r=1}^{N_{f}-N_{c}}\bar{{\tilde{\nu}}}_{\dot{\alpha}r}
(\bar{\sigma}^{a})^{\dot{\alpha}\alpha}{\tilde{\nu}}_{\alpha}^{r}
\eea
where, in the second line, we have invoked the decomposition
\eqn{decomp} such that the fermi bilinear $\hat{\Lambda}$ does not
contain any super-orientation modes $\nu$ or $\tilde{\nu}$,
\be \hat{\Lambda}_a = \bar{\chi}_{\dot{\alpha}}
(\bar{\sigma}^{a})^{\dot{\alpha}\alpha} \chi_{\alpha}
+\chi^{\alpha}(\sigma^{a})_{\alpha\dot{\alpha}}
\bar{\chi}^{\dot{\alpha}}-
\theta^{\alpha}(\sigma^{a})_{\alpha\dot{\alpha}}\,W
\bar{\theta}^{\dot{\alpha}} -
\bar{{\tilde{\theta}}}_{\dot{\alpha}}
(\bar{\sigma}^{a})^{\dot{\alpha}\alpha}
\,\tilde{W}{\tilde{\theta}}_{\alpha} \label{hatlam}\ee
Integrating out the $X$ fields gives schematically ${\mathbb
L}X\sim \Lambda$ which, upon substitution back into the action,
yields the appropriate four-fermi terms which may saturate the
integration over $\nu$. However, one of the key insights of the
analysis of \cite{dhkmv} is that it pays dividends to keep the $X$
fields rather than to write them in terms of fermi bi-linears
$\Lambda$. Indeed, performing the integration over the $k(N_c-k)$
variables $\nu$ and $k(N_f-N_c-k)$ variables $\tilde{\nu}$ yields
\bea \int\,d\nu \,d\tilde{\nu}
\,\exp\left( -{\rm Tr}_{k}\, \sum_{i=1}^{N_c}\nu^{\alpha}_i
X_{\alpha\dot{\alpha}}\bar{\nu}^{\dot{\alpha} i}
-{\rm Tr}_k\sum_{r=1}^{N_f-N_c}\bar{\tilde{\nu}}^{\dot{\alpha}}_r
\bar{X}^{\dot{\alpha}\alpha}\tilde{\nu}^r_{\alpha}\right) =({\rm
det}_{2k}\, X)^{N_{f}-2k}\nn\eea
with $X_{\alpha\dot{\alpha}}=X^a(\sigma_a)_{\alpha\dot{\alpha}}$
and
$\bar{X}^{\dot{\alpha}\alpha}=X^a(\bar{\sigma}_a)^{\dot{\alpha}\alpha}$,
each a $2k\times 2k$ matrix. Putting everything together, we have
an expression for the effective partition function for $k$
instantons,
\begin{eqnarray} {\mathcal Z}_{k,N_{f},N_{c}} &=&
\frac{C_{k,N_c}C_{k,N_f-N_c}}{{\rm Vol}(U(k))} \int\,dX\, da\,
dW\, d\tilde{W}\, d\chi \,d\theta\,d\tilde{\theta}\ [\Delta]\
({\rm det}_{2k}\,X_a)^{N_{f}-2k} \nn\\ && \ \ \ \ \times\left({\rm
det}_{k} W\right)^{N_{c}-2k}({\rm
det}_{k}\tilde{W})^{(N_{f}-N_{c})-2k}
\exp\left(-\,{\rm Tr}_k(X_a{\mathbb L}[X]+X_a\hat{\Lambda}^a)\right)\,.\nonumber\\
\nn\end{eqnarray}
where the normalization coefficients $C_{k,N}$, the linear
operator ${\mathbb L}$ and the fermi bi-linear $\hat{\Lambda}$ are
defined in \eqn{ckn}, \eqn{L} and \eqn{hatlam} respectively. The
constraints \eqn{higgs} and \eqn{newfcons} are imposed by the
Dirac delta functions, indicated by $[\Delta]$.
%
%
However, as noted, previously, these constraints are linear in our
variables $W, \tilde{W}$ allowing us to perform the integrals. We
make the trivial change of variables,
\be W_\pm = W\pm\tilde{W}\ee
so that the bosonic constraint \eqn{higgs} reads simply
$W_-=[a,a^\dagger]$ and is removed by integration over $W_-$. Indeed, 
if we are interested in operators of the 2d SCFT which are $U(N_c)\times 
SU(N_f)$ singlets (as opposed to merely $U(N_c)\times U(N_f-N_c)$ singlets 
\eqn{breaking}) then they may only depend on $W_+$ in the instanton background. 
We mimic 
this procedure for the fermions, defining $\theta_\pm=\theta\pm\tilde{\theta}$. 
The fermionic constraints \eqn{newfcons} remove one half of our
fermionic integrations which we take to be $\theta_-$ and $\bar{\theta}_-$, 
picking up further powers of $\det^kW$ and ${\rm det}^k\,\tilde{W}$. This 
leaves us with our final, unconstrained result, for the instanton partition
function
\bea {\cal
Z}_{k,N_c,N_f}&=&\frac{C_{k,N_c}C_{k,N_f-N_c}}
{2^{kN_{f}-k^2}\,{\rm
Vol}(U(k))} \int\,dX\,da\,dW_+\,d\chi\,d\theta_+\
\left({\rm det}_{2k}\,X\right)^{N_f-2k}\,\left({\rm det}_{k}(W_+
+[a,a^\dagger])\right)^{N_c-k}\nn\\ &&\ \ \ \ \ \ \times
\left({\rm det}_{k} ({W}_+-[a,a^\dagger]) \right)^{N_f-N_c-k}
\exp\left(-{\rm Tr}_k (X_a\mathbb{L}[X]+X_a\hat{\Lambda}^a)\right)
\nn\eea
where we remind the reader that all integrations are over the variables and 
their complex conjugates.

\subsubsection*{The Partition Function for a Single Instanton}

In the case of a single $k=1$ instanton, the partition function
simplifies tremendously, aided by the fact that the adjoint chiral
multiplet fields $a$ and $\chi$ decouple. The scalar $a=-\sqrt{2\pi r}z$ is
simply the position of the instanton, while the complex fermion
$\chi^\alpha$, which can be rewritten as four real Grassmann
parameters $\xi^a$, are the Goldstino modes of the instanton
arising broken ${\cal N}=(4,4)$ supersymmetry.

\para
The hypermultiplet field $W_+$ is related to the scale size of the instanton
\be W_+=\rho^2 \ee
where the exponent follows from noting the scaling dimension $[W_+]=-2$. The 
fermionic superpartners living in $\theta_+$ are 
identified with the four superconformal modes of the instanton which we 
shall denote as $\eta^a$. The $X$ determinant is trivially  evaluated ${\rm det_2
X}=X_aX_a$. Collecting everything together, we have
\bea {\mathcal{Z}}_{1,N_{f},N_{c}}
&=&\frac{2r\,\pi^{N_{f}}N_c(N_f-N_c)}{N_{c}!(N_{f}-N_{c})!}\int\,d^2z\,d^4\xi\,d^4\eta
\nn\\ &\times&\int\, d(\rho^2)\,d^{4}X \
(X_aX_a)^{N_f-2}\,\rho^{2N_f-4}\,\exp(-\rho^2 X_aX_a) \eea
The integral in the final line may be trivially evaluated. We
split the $X_a$, $a=1,2,3,4$ into radial and polar parts
$(u,\hat{\Omega}_3)$. Since we wish to use our effective partition 
function to compute correlation functions with $Spin(4)_R$ charge, 
we integrate only over the radial direction while leaving the 
3-sphere $\hat{\Omega}_3$ intact.
(Indeed, in the $N_f\rightarrow\infty$ limit a saddle
point approximation restricts the integral to $u\sim \rho^{-1}$). 
Our final result for the partition
function for a single instantons is
\be {\mathcal{Z}}_{1,N_{f},N_{c}}=\,
\frac{2r\,\pi^{N_f}(N_f-1)!}{(N_c-1)!(N_f-N_c-1)!}\
\int\,d^2z\frac{d\rho}{\rho^{3}}\,d\hat{\Omega}_{3}\ d^4\xi
d^4\eta \label{atlast}\ee
which indeed reveals the $AdS_3\times S^3$ structure as promised.
This partition function may be used to compute the instanton 
contribution to any suitable gauge and flavor
singlet correlation function which saturates the 8 supersymmetry
and superconformal fermi zero modes.

\para
{\bf Acknowledgements}: We are especially grateful to  Koji
Hashimoto for numerous patient and insightful discussions. We
would also like to thank Nick Dorey, Yang-Hui He, Eva Silverstein
and Aninda Sinha for useful comments. HYC is supported by
St.John's college, Cambridge through a Benefactors Scholarship. DT
is supported by the Royal Society.


\begin{thebibliography}{99}

\bibitem{ww}
 S.~Weinberg and E.~Witten,
  ``{\it Limits On Massless Particles},''
  Phys.\ Lett.\ B {\bf 96}, 59 (1980).

\bibitem{juan}
  J.~M.~Maldacena,
``{\it The large $N$ limit of superconformal field theories and
supergravity},''
  Adv.\ Theor.\ Math.\ Phys.\  {\bf 2}, 231 (1998)
  [Int.\ J.\ Theor.\ Phys.\  {\bf 38}, 1113 (1999)]
  [arXiv:hep-th/9711200].
%
\bibitem{greenbank}  T.~Banks and M.~B.~Green,
  ``{\it Non-perturbative effects in $AdS_5\times S^5$ string theory and $d=4$ SUSY
  Yang-Mills},''
  JHEP {\bf 9805}, 002 (1998)
  [arXiv:hep-th/9804170].

\bibitem{colour}
M.~Bianchi, M.~B.~Green, S.~Kovacs and G.~Rossi,
  ``{\it Instantons in supersymmetric Yang-Mills and D-instantons in IIB
  superstring theory},''
  JHEP {\bf 9808}, 013 (1998)
  [arXiv:hep-th/9807033].

\bibitem{dhkmv}
N.~Dorey, T.~J.~Hollowood, V.~V.~Khoze, M.~P.~Mattis and
S.~Vandoren, ``{\it Multi Instantons and Maldacena's
conjecture},'' JHEP {\bf 9906}, 023 (1999)
  [arXiv:hep-th/9810243];
``{\it Multi Instanton calculus and the AdS/CFT correspondence in
${\cal N}=4$ superconformal field theory},'' Nucl.\ Phys.\ B {\bf
552} (1999) 88 [arXiv:hep-th/9901128].



\bibitem{n2}
 T.~J.~Hollowood, V.~V.~Khoze and M.~P.~Mattis,
  ``{\it Summing the instanton series in ${\cal N}=2$ superconformal large-N QCD},''
  JHEP {\bf 9910}, 019 (1999)
  [arXiv:hep-th/9905209].

\bibitem{orbi}
T.~J.~Hollowood and V.~V.~Khoze,
  ``{\it ADHM and D-instantons in orbifold AdS/CFT duality},''
  Nucl.\ Phys.\ B {\bf 575}, 78 (2000)
  [arXiv:hep-th/9908035].

\bibitem{spn1}
 E.~Gava, K.~S.~Narain and M.~H.~Sarmadi,
  ``{\it Instantons in ${\cal N}=2$ $Sp(N)$ superconformal gauge theories and the AdS/CFT
  correspondence},''
  Nucl.\ Phys.\ B {\bf 569}, 183 (2000)
  [arXiv:hep-th/9908125].

\bibitem{spn}
 T.~J.~Hollowood,
  ``{\it Instantons, finite ${\cal N}=2$ $Sp(N)$ theories and the AdS/CFT correspondence},''
  JHEP {\bf 9911}, 012 (1999)
  [arXiv:hep-th/9908201];

  \bibitem{spn2} T.~J.~Hollowood, V.~V.~Khoze and M.~P.~Mattis,
  ``{\it Instantons in ${\cal N}=4$ $Sp(N)$ and $SO(N)$ theories and the AdS/CFT
  correspondence},''
  Adv.\ Theor.\ Math.\ Phys.\  {\bf 4}, 545 (2000)
  [arXiv:hep-th/9910118].


\bibitem{coulomb}
 M.~B.~Green and C.~Stahn,
  ``{\it D3-branes on the Coulomb branch and instantons},''
  JHEP {\bf 0309}, 052 (2003)
  [arXiv:hep-th/0308061].

\bibitem{beta}
 G.~Georgiou and V.~V.~Khoze,
  ``{\it Instanton calculations in the beta-deformed AdS/CFT correspondence},''
  arXiv:hep-th/0602141.

\bibitem{brownh}
J.~D.~Brown and M.~Henneaux,
  ``{\it Central Charges In The Canonical Realization Of Asymptotic Symmetries: An
  Example From Three-Dimensional Gravity},''
  Commun.\ Math.\ Phys.\  {\bf 104}, 207 (1986).


\bibitem{review}
O.~Aharony, S.~S.~Gubser, J.~M.~Maldacena, H.~Ooguri and Y.~Oz,
``{\it Large N field theories, string theory and gravity},''
  Phys.\ Rept.\  {\bf 323}, 183 (2000)
  [arXiv:hep-th/9905111].


\bibitem{mik}
A.~Mikhailov,
  ``{\it D1D5 system and noncommutative geometry},''
  Nucl.\ Phys.\ B {\bf 584}, 545 (2000)
  [arXiv:hep-th/9910126].


\bibitem{amime}
  A.~Hanany and D.~Tong,
  ``{\it Vortices, instantons and branes},''
  JHEP {\bf 0307} (2003) 037
  [arXiv:hep-th/0306150].

\bibitem{diacseib}
D.~E.~Diaconescu and N.~Seiberg,
  ``{\it The Coulomb branch of (4,4) supersymmetric field theories in two
  dimensions},''
  JHEP {\bf 9707}, 001 (1997)
  [arXiv:hep-th/9707158].

\bibitem{wittenhiggs}
E.~Witten, ``{\it On the conformal field theory of the Higgs
branch},''
  JHEP {\bf 9707}, 003 (1997)
  [arXiv:hep-th/9707093].




\bibitem{ofer}
 O.~Aharony and M.~Berkooz,
  ``{\it IR dynamics of $d=2$, ${\cal N}=(4,4)$ gauge theories and DLCQ of 'little  string
  theories},''
  JHEP {\bf 9910}, 030 (1999)
  [arXiv:hep-th/9909101].


\bibitem{swd1d5}
  N.~Seiberg and E.~Witten,
  ``{\it The D1/D5 system and singular CFT},''
  JHEP {\bf 9904} (1999) 017
  [arXiv:hep-th/9903224].

\bibitem{brodie}
 J.~H.~Brodie,
  ``{\it Two dimensional mirror symmetry from M-theory},''
  Nucl.\ Phys.\ B {\bf 517}, 36 (1998)
  [arXiv:hep-th/9709228].

\bibitem{mohsen}
 M.~Alishahiha,
  ``{\it N=(4,4) 2D supersymmetric gauge theory and brane configuration},''
  Phys.\ Lett.\ B {\bf 420}, 51 (1998)
  [arXiv:hep-th/9710020];
  ``{\it On the brane configuration of N=(4,4) 2D supersymmetric gauge
  theories},''
  Nucl.\ Phys.\ B {\bf 528}, 171 (1998)
  [arXiv:hep-th/9802151].




\bibitem{wittphases}
 E.~Witten,
  ``{\it Phases of N = 2 theories in two dimensions},''
  Nucl.\ Phys.\ B {\bf 403}, 159 (1993)
  [arXiv:hep-th/9301042].

\bibitem{semi}  T.~Vachaspati and A.~Achucarro,
  ``{\it Semilocal cosmic strings},''
  Phys.\ Rev.\ D {\bf 44}, 3067 (1991).

\bibitem{schroers}
B.~J.~Schroers,
  ``{\it The Spectrum of Bogomol'nyi Solitons in Gauged Linear Sigma Models},''
  Nucl.\ Phys.\ B {\bf 475}, 440 (1996)
  [arXiv:hep-th/9603101].

\bibitem{auzzi}
  R.~Auzzi, S.~Bolognesi, J.~Evslin, K.~Konishi and A.~Yung,
 ``{\it Nonabelian superconductors: Vortices and confinement in N = 2 SQCD},''
  Nucl.\ Phys.\ B {\bf 673}, 187 (2003)
  [arXiv:hep-th/0307287].

\bibitem{ward}
  R.~S.~Ward,
  ``{\it Slowly Moving Lumps In The $CP^1$ Model In (2+1)-Dimensions},''
  Phys.\ Lett.\ B {\bf 158}, 424 (1985).

\bibitem{leese}  R.~A.~Leese and T.~M.~Samols,
  ``{\it Interaction of semilocal vortices},''
  Nucl.\ Phys.\ B {\bf 396}, 639 (1993).

\bibitem{sy}
 M.~Shifman and A.~Yung,
 ``{\it Non-Abelian semilocal strings in N = 2 supersymmetric QCD},''
  arXiv:hep-th/0603134.

\bibitem{vstring}
  A.~Hanany and D.~Tong,
  `{\it Vortex strings and four-dimensional gauge dynamics},''
  JHEP {\bf 0404}, 066 (2004)
  [arXiv:hep-th/0403158].

\bibitem{tit} 
M.~Eto, Y.~Isozumi, M.~Nitta, K.~Ohashi and N.~Sakai,
  ``{\it Moduli space of non-Abelian vortices},''
  arXiv:hep-th/0511088.

\end{thebibliography}
\end{document}